# Quantitative depth profiling of $Si_{1-x}Ge_x$ structures by time-of-flight secondary ion mass spectrometry and secondary neutral mass spectrometry

M.N. Drozdov [a,b], Y.N. Drozdov [a,b], A. Csik [c], A.V. Novikov [a,b], K. Vad [c], P.A. Yunin [a,b], D.V. Yurasov [a,b], S.F. Belykh [d], G.P. Gololobov [e], D.V. Suvorov [e], A. Tolstogouzov [e,f]

[a] *Institute for Physics of Microstructures of the Russian Academy of Sciences (IPM RAS), 603950 Nizhniy Novgorod, Russian Federation*
[b] *Lobachevski Nizhniy Novgorod State University, 603950 Nizhniy Novgorod, Russian Federation*
[c] *Institute for Nuclear Research (INR), Hungarian Academy of Science, Bem tér 18/C, 4026 Debrecen, Hungary*
[d] *MATI Russian State Technological University, Orshanskaya Str. 3, 121552 Moscow, Russian Federation*
[e] *Ryazan State Radio Engineering University, Gagarin Str. 59/1, 390005 Ryazan, Russian Federation*
[f] *Centre for Physics and Technological Research (CeFITec), Dept. de Física da Faculdade de Ciências e Tecnologia (FCT), Universidade Nova de Lisboa, 2829-516 Caparica, Portugal*

**ABSTRACT**

Quantification of Ge in $Si_{1-x}Ge_x$ structures ($0.092 \leq x \leq 0.78$) was carried out by time-of-flight secondary ion mass spectrometry (TOF-SIMS) and electron-gas secondary neutral mass spectrometry (SNMS). A good linear correlation ($R^2 > 0.9997$) of the intensity ratios of secondary ions $GeCs_2^+/SiCs_2^+$ and $^{74}Ge^-/^{30}Si^-$ and post-ionized sputtered neutrals $^{70}Ge^+/^{28}Si^+$ with Ge concentration was obtained. The calibration data were used for quantitative depth profiling of [10 × (12.3 nm $Si_{0.63}Ge_{0.37}$/34 nm Si)] structures on Si. Satisfactory compliance of the quantified Ge concentration in SiGe layers with the values obtained by high-resolution X-ray diffraction was revealed for both techniques. SIMS and SNMS experimental profiles were fitted using Hofmann's mixing-roughness-information depth (MRI) model. In the case of TOF-SIMS, the quality of the reconstruction was better than for SNMS since not only the progressing roughening, but also the crater effect and other processes unaccounted in the MRI simulation could have a significant impact on plasma sputter depth profiling.

## 1. Introduction

An unceasing progress in silicon–germanium technology (see, e.g., [1,2] and references cited therein) stimulates interest in quantification of SiGe alloys and multilayer structures by different analytical techniques, especially by secondary ion mass spectrometry (SIMS). In the late eighties, Gillen et al. [3] using $Ar^+$, $O_2^+$ and $Cs^+$ primary ions revealed considerable variations in Si and Ge secondary ion yields preventing quantification of the sputter depth profiles. However, many subsequent studies [4–22] (to name some but not all works) have demonstrated the possibility in principle to quantify $Si_{1-x}Ge_x$ systems over a wide range of germanium concentration.

Relative sensitivity factor (RSF) [23] is usually employed for SIMS quantification of diluted samples, and atom density of impurity $\rho_i$ (at/cm$^3$) can be estimated as

$$\rho_i = \frac{I_i}{I_m} RSF_i, \quad (1)$$

where $I_i$ and $I_m$ are the isotope secondary ion intensities of impurity and matrix elements, respectively, and $RSF_i$ is the relative sensitivity factor for a given impurity in a given matrix estimated for given experimental conditions (type of instrument, primary ions, secondary ions, etc.).

The tabulated RSF data [23,24] have units of at/cm$^3$, and the relative sensitivity factor for a matrix element is equal to its atom density $\rho_m$ (for example, $RSF_{Si} = 4.99 \times 10^{22}$ at/cm$^3$). A single implanted standard is sufficient for calibration, and a conventional RSF approach presupposes the linear dependence of a SIMS signal of a given impurity on its concentration in a given matrix. Such an approach has proved itself for a trace level of impurities ($\rho_i < 1$ at.%), when the matrix atom density is considered to be constant.

To quantify germanium in $Si_{1-x}Ge_x$ systems, the conventional *RSF* approach has been modified, and Eq. (1) takes the form

$$\frac{I_{Ge}}{I_{Si}} = k \frac{x}{1-x}, \qquad (2)$$

where $I_{Ge}$ and $I_{Si}$ are the intensities of atomic or polyatomic secondary ion species of germanium and silicon, positively or negatively charged, *x* is the atomic fraction of germanium, and *k*-factor can be considered as a ratio of relative sensitivity factors for Si and Ge

$$k = \frac{RSF_{Si}}{RSF_{Ge}}. \qquad (3)$$

The success of this strategy requires constancy of *k*-factor under the changing of Ge and, correspondingly, Si atomic fraction in SiGe systems. This condition is not only limited by linear dependence of Ge and Si secondary ion intensities on the concentration of these elements (constant *RSF*s, as for diluted samples), but also assumes correlated variation of relative sensitivity factors due to a similar influence (counterbalancing) of matrix effects on the Ge and Si secondary ion yields. Mainly, matrix effects depend on the electronic and vibrational states of both sputtered species as well as on the chemical bonding of these species to the surface. Taking into account a similar electronic configuration of IV-group elements, the compensation of matrix effects in SiGe systems is permissible. This has been confirmed using different combinations of primary and secondary ions. The numerical value of *k*-factor depends on the type and polarity of these ions and instrumental realization. Several standards with different *x*-values are required to achieve high calibration accuracy.

At present, the most widespread SIMS quantitative approach employs a $Cs^+$ primary ion beam with registration of $MCs^+$ secondary molecular ions (where M denotes the element to be analyzed). This approach was firstly applied for $A^{III}B^V$ compound semiconductors [25], and then has been successfully tested for $Si/Si_{1-x}Ge_x/Si$ heterostructures with *x*-value varying from 0 to 0.235 [5]. The constancy of *k*-factor for $GeCs^+/SiCs^+$ intensity ratio was obtained, in spite of the non-linear dependence of these ion signals on Ge concentration. In addition, the constant *k*-factor was found for $^{70}Ge^+/^{30}Si^+$ ratio with $O_2^+$ primary ion bombardment. The measurements were carried out using a sector magnetic instrument Cameca IMS 4f.

In the past fifteen years, systematic studies of $Si_{1-x}Ge_x$ systems have been continued for a wider range of Ge and for different experimental conditions. For example, matrix effects were compensated for $0.07 \leq x \leq 0.27$ by measuring $^{70}Ge^-/^{30}Si^-$ intensity ratio with $Cs^+$ primary ion bombardment [8], and possibility to use for quantification the intense $GeCs_2^+$ and $SiCs_2^+$ secondary molecular ions was shown in [11]. However, later on these results were not confirmed.

Many studies were carried out using $O_2^+$ primary ions [4,6,7,9–13,15] including isotope $^{18}O_2^+$ beam [16]. Constant *k*-factor was revealed for $Ge^+/Si^+$ intensity ratio with low-energy $O_2^+$ bombardment for $0.15 \leq x \leq 0.65$ [9]. However, in other research [10] tremendous variation of this ratio was found for $0.3 \leq x \leq 0.65$. The sputtering by low-energy oxygen ions results in the formation of an altered layer and strong surface roughening, which depends on an impact energy and the angle of incidence of primary ions. These artifacts complicate calibration procedure, but did not impede quantification of $Si_{1-x}Ge_x$ multilayers including their interfaces [15,16]. There are also some investigations with inert gas ($Ar^+$, $Kr^+$) primary ion beams [12,14]. However, in that case the secondary ion yields are smaller than in the case of $Cs^+$ and $O_2^+$ ion beam bombardment.

The studies mentioned above have been carried out using sector magnetic instruments Cameca IMS 3f–5f, 7f, Wf and quadrupole-based SIMS, mainly Atomica 4500. At present, time-of-flight secondary ion mass spectrometry (TOF-SIMS) is employed in semiconductor technology too. In contrast to magnetic and quadrupole-based SIMS, where the same primary ion beam is used for controlled sputtering of a sample and for generation of analytical ion signals, TOF-SIMS depth profiling is performed using different ion beams for sputtering and probing (dual beam mode). Comparative study of $Si_{1-x}Ge_x$ ($0 < x \leq 0.85$) with Cameca IMS-5f and TOF.SIMS-5 by IONTOF was carried out using the $MCs^+$ approach with a 2 keV $Cs^+$ ion beam [17]. A good linear correlation of the secondary ion intensity ratio $GeCs_n^+/SiCs_n^+$, where n = 1, 2, with Ge concentration was revealed for both instruments. However, the linear fit was slightly less good for TOF.SIMS-5 as compared with Cameca IMS-5f, namely, the correlation coefficient $R^2$ was 0.9998 for $MCs^+$ and 0.9999 for $MCs_2^+$ intensity ratios in the case of a magnetic instrument, and 0.9997 and 0.9974, respectively, for the time-of-flight spectrometer.

It was found [18] that sputtering by $Cs^+$ ions is more preferable for depth profiling of $Ge_xSi_{1-x}/Si$ heterostructures than by $O_2^+$ ions, since $Cs^+$ ions have allowed minimizing ion-induced surface roughening. The experiments were carried out by a TOF.SIMS-5 with the bombarding energies in the range from 0.5 to 2 keV and the same angle of incidence (45°) of both sputter ion beams. While the root-mean-square roughness $\sigma$ increased only slightly for $Cs^+$ ions, from 0.8 to 1.2 nm, it reached 3–5 nm after $O_2^+$ sputtering at the depth of 1 µm.

Recently, Py et al. using a TOF.SIMS-5 have shown [19,20] that in the case of 1 keV $Cs^+$ sputtering of $Si_{1-x}Ge_x$ alloys $^{70}Ge^-/^{30}Si^-$ intensity ratio remains constant only for $x \leq 0.33$ and for greater value of Ge concentration non-linear behavior was observed, hampering the precision of quantification. An alternative protocol was tested, the full-spectrum method [26], which states the proportionality between total intensities of the secondary ion peaks and composition of the actual material. $Si_{1-x}Ge_x$ layers were depth profiled with a long cycle time allowing the detection of $Si_nGe_m$ polyatomic secondary ions (n, m = 1–6), and compensation of matrix effects were observed within the whole range of *x*-values. It was concluded that application of the full-spectrum method is equivalent to the measuring of sputtered neutral fraction, for which matrix effects are negligible. However, from the practical point of view this approach is labor consuming since it requires prolonged monitoring of the intensities of at least 40 mass peaks during of the depth profiling of $Si_{1-x}Ge_x$ systems.

Secondary ions constitute a small fraction of sputtered species, and registration of neutrals using post-ionization techniques looks preferable for correct quantification of matrix elements. To the best of our knowledge, the only comparative study of $Si_{1-x}Ge_x$ samples by SIMS and the electron-gas version of secondary neutral mass spectrometry (hereinafter referred as SNMS) was reported in the early nineties [27]. In this technique [28], the post-ionizing electrons are provided by the electron component of low-pressure noble gas plasma (mostly, in argon) being excited by electron cyclotron wave resonance, and plasma ions are extracted onto a sample surface to perform controlled erosion. In [27], SIMS depth profiling of a-$Si_{1-x}Ge_x$:H samples ($0 < x \leq 0.58$) was performed with a Cameca IMS-4f using a 5.5 keV $Cs^+$ primary ion beam, and SNMS measurements were carried out with an INA-3 by Leybold [29]. SNMS was used as a certificated method for quantification of Ge concentration. The $Ge^+/Si^+$ signal ratio was equal to 1.6, but accuracy and other details of the calibration procedure were not discussed.

In the present study, we performed comparative quantitative analysis of $Si_{1-x}Ge_x$ ($0.092 \leq x \leq 0.78$) structures by time-of-flight secondary mass spectrometry and by electron-gas secondary neutral mass spectrometry. A TOF.SIMS-5 and INA-X by SPECS [30], an advanced version of INA-3, have been involved in our experiments. This work is a continuation and further development of our previous SIMS research of $Si_{1-x}Ge_x$ structures [21,22], in which germanium atomic fraction did not exceed 0.6. The calibration SIMS data, collecting for molecular secondary ions $GeCs_2^+$ and $SiCs_2^+$, and for atomic $Ge^-$ and $Si^-$ ions were used for quantitative depth profiling of [10 × (12.3 nm $Si_{0.63}Ge_{0.37}$/34 nm Si)] stacks deposited on Si. A similar research was carried out by SNMS, but employing for calibration the intensity ratio of post-ionized Ge and Si sputtered neutrals. To understand an influence of bombardment-induced effects on the quantification of experimental SIMS and SNMS depth profiles, we reconstructed initial Ge in-depth distributions using Hofmann's mixing-roughness-information depth (MRI) model [31].

## 2. Experimental details

Three reference samples were prepared by molecular-beam epitaxy using a SIVA-21 by Riber at IPM RAS. The $Si_{1-x}Ge_x$ structures on Si denoted as S1 and S2 samples consisted of three 200-nm thick layers of germanium–silicon solid solutions with $x$-values varying in the range from 0.092 to 0.583. High-resolution X-ray diffractometry (HR-XRD) using a D8 Discover by Bruker was employed for estimation of Ge atomic fraction in these layers taking into account the deviation from Vegard's law. The values of Ge concentration and the degree of elastic stress relaxation of each layer in the S1 and S2 samples are presented in Table 1. Additionally, a 500-nm thick $Ge_{0.78}Si_{0.22}$ structure (S3 sample) grown on a Ge (001) substrate was used for calibration. Verification and comparison of the different calibration approaches using the positive molecular and negative atomic ions were made by quantitative depth profiling of the [10 × (12.3 nm $Si_{0.63}Ge_{0.37}$/34 nm Si)] structure covered by a 150 nm-thick Si layer (S4 sample).

TOF-SIMS calibration and depth profiling were carried out at IPM RAS using a time-of-flight secondary ion mass spectrometer TOF.SIMS-5. The instrument operates in dual beam mode employing 1 keV/80 nA $Cs^+$ ions for sputtering with a 45° angle of incidence. This beam was scanned over an area from 200 × 200 μm² to 400 × 400 μm², and the analyzed region was ca. 4% in square around the center of the sputter crater. The pulsed 25 keV/1 pA $Bi^+$ and $Bi_3^+$ beams were used for probing with the same angle of incidence as in the case of the sputter beam. For analysis we selected positive molecular ions $GeCs_n^+$ and $SiCs_n^+$, where n = 1, 2, and negative atomic ions $Ge^-$ and $Si^-$. The crater depths were measured by an optical interference microscope Talysurf CCI-2000. Detailed description of the experimental conditions can be found elsewhere [18].

SNMS measurements were carried out in an electron-gas secondary neutral mass spectrometer INA-X at INR HAS. In the direct bombardment mode, $Ar^+$ ions are extracted from low-pressure plasma and bombarded a negatively biased (−277 V) sample surface with a current density of ca. 1 mA/cm². The sputtered area was confined to a circle of 2 mm in diameter by a Ta mask. Leaving the plasma, post-ionized neutrals are directed into a quadrupole mass filter Balzers QMA 410 by means of electrostatic lenses and a broad-pass energy analyzer. As analyzed ions, positively charged $Ge^+$ and $Si^+$, are employed. The sputtering rates were calculated from the crater depth measured by an AMBIOS XP-1 stylus-type profilometer.

## 3. Results and discussion

### 3.1. TOF-SIMS calibration

Depth profiles of the positive molecular secondary ions collected for the reference sample S1 are shown in Fig. 1a. A cluster probing $Bi_3^+$ beam was employed since it provides secondary ion yields 3–5 times greater as compared with an atomic $Bi^+$ beam. The intensities of $GeCs_2^+$ and $SiCs_2^+$ secondary ions were higher than $GeCs^+$ and $SiCs^+$, as it was also found in [11], and $^{74}GeCs_2^+$ and $^{28}SiCs_2^+$ ions were used for calibration in our study. Similar profiles were measured for the reference samples S2 and S3. After averaging the data for each $Si_{1-x}Ge_x$ layer, the calibration dependence was obtained (Fig. 1b). The correlation coefficient $R^2$ of the linear fit was found to be 0.9999, better than that obtained in [17] using analogous TOF.SIMS-5 instrument.

**Table 1**
Ge concentration and the degree of elastic stress relaxation in the reference samples.

| Sample | Layer's number[a] | Ge concentration, at.% | Degree of relaxation |
|---|---|---|---|
| S1 | 1 | 9.2 ± 0.5 | 0.08 |
|  | 2 | 29.2 ± 0.5 | 0.75 |
|  | 3 | 47.2 ± 1 | 0.76 |
| S2 | 1 | 21.5 ± 0.5 | 0.8 |
|  | 2 | 40.8 ± 0.5 | 0.9 |
|  | 3 | 58.3 ± 0.5 | 0.8 |

[a] Layer's numeration from Si substrate towards the surface.

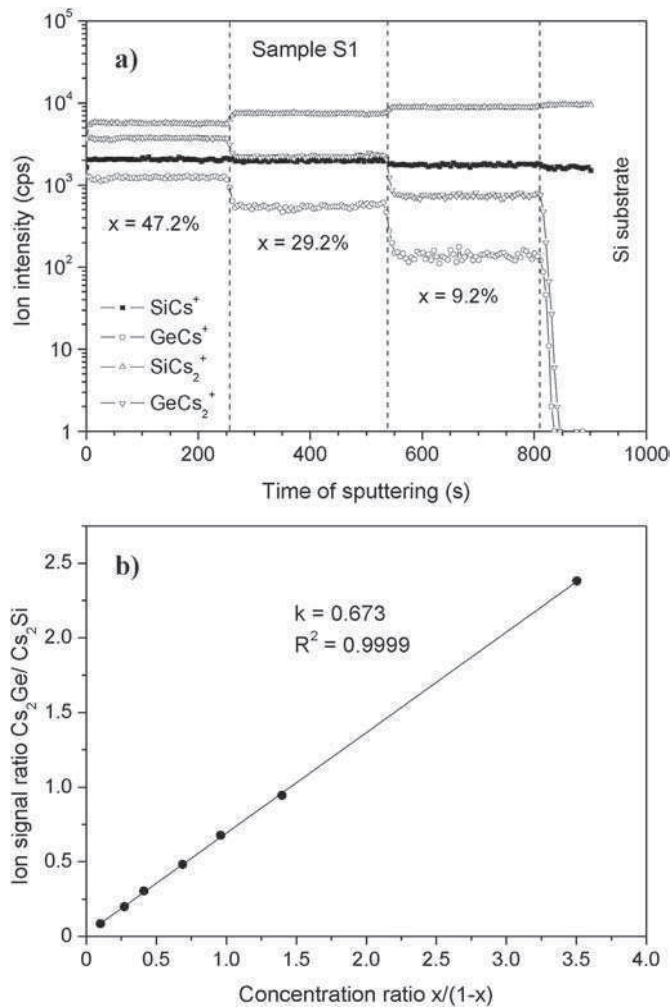

**Fig. 1.** TOF-SIMS depth profiles of the positive molecular secondary ions measured for sample S1 with $Bi_3^+$ probing ions (a), and linear fit of the calibration dependence obtained for samples S1–S3 (b), where $R^2$ is the correlation coefficient and $k$ is the calibration factor from Eqs. 2 and 3.

Depth profiles of the negative atomic secondary ions collected for the S1 sample are presented in Fig. 2a. Similar results were obtained for the S2 and S3 samples. Atomic $Bi^+$ probing ions were used, and the intensity of the most abundant $^{74}Ge$ isotopic ions was measured. In most studies, the peak of the less abundant $^{70}Ge$ isotope was monitored since in the case of the negative secondary ion measurement $^{74}Ge$ peak is overlapped by intense $^{29}Si_2O$ and $^{28}Si^{30}SiO$ peaks. High mass resolution ($M/\Delta M > 10^4$) realized in a TOF.SIMS-5 allowed avoiding undesirable mass spectral interference.

Fig. 2b shows the calibration dependence obtained for the negatively charged atomic secondary ions. The linear fit was also good ($R^2 = 0.9997$) within the whole range of Ge concentration, from 9.2 to 78 at.%. That is better than in [19], where considerable deviation from the linear dependence was observed at $x > 0.33$.

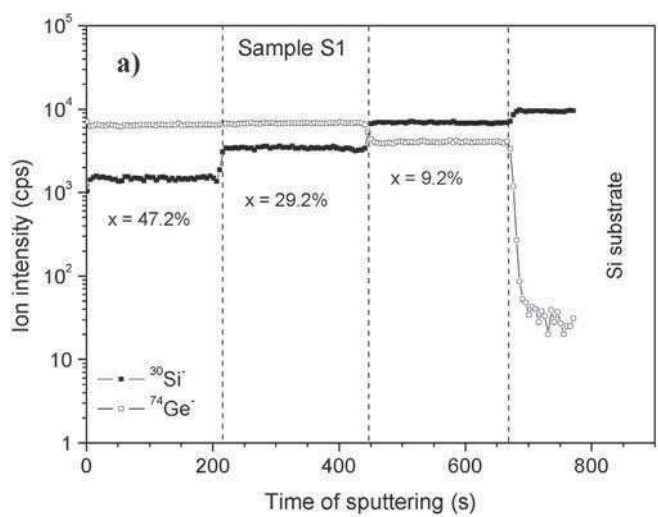

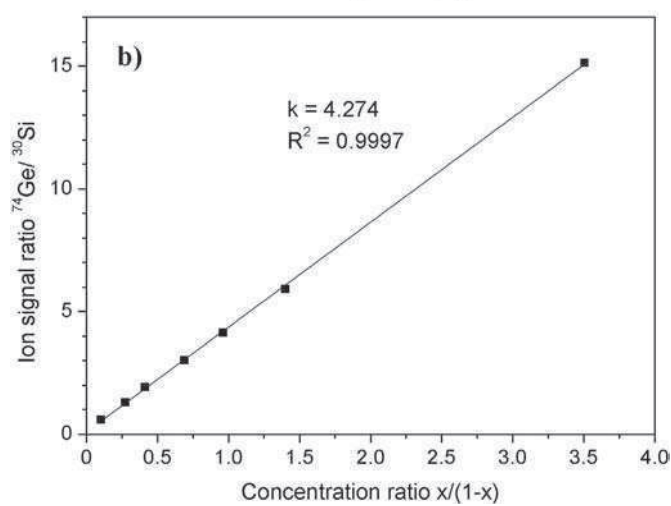

**Fig. 2.** TOF-SIMS depth profiles of the negative atomic secondary ions measured for sample S1 with $Bi^+$ probing ions (a), and linear fit of the calibration dependence obtained for samples S1–S3 (b), where $R^2$ is the correlation coefficient and $k$ is the calibration factor from Eqs. 2 and 3.

### 3.2. SNMS calibration

Representative mass spectrum of the post-ionized sputtered neutrals measured for the sample S3 ($Ge_{0.78}Si_{0.22}$) is presented in Fig. 3. Besides the peaks of Si and Ge atomic and polyatomic ions, one can see (i) doubly charged Ta ions from a mask, (ii) Ar ions, including doubly charged ones, originated from plasma (a very intense $^{40}Ar^+$ peak was automatically suppressed during mass spectra acquisition), and (iii) the peaks of common surface contaminants (C, N and O).

Depth profiles of the post-ionized Ge and Si sputtered atoms measured for the sample S1 are presented in Fig. 4a, and the calibration dependence for the samples S1–S3 with the proper linear fit is shown in Fig. 4b. The quality of approximation ($R^2 = 0.9999$) is very good, as in the case of TOF-SIMS measurements of the positive molecular secondary ions. Since INA-X is equipped with a quadrupole mass filter, we monitored the intensity of $^{70}Ge$ isotopic ions that allowed decreasing an influence of mass spectral interference. The $k$-factor for $^{70}Ge^+/^{28}Si^+$ ratio was found to be 0.2777 that gives a value of 1.21 for $Ge^+/Si^+$ signal ratio. That is slightly lower than the number obtained for a-$Si_{1-x}Ge_x$:H heterostructures using an INA-3 instrument [27].

In our study, the calibration was made for $0.092 \leq x \leq 0.78$, and the linear fit of the experimental Ge/Si signal ratio shown in Figs. 1b, 2b, 4b confirms the constant yield ratio assumption for the whole concentration range of Ge, both for the secondary ions (positive molecular and negative atomic) and post-ionized sputtered neutrals. However, a small but finite intercept for all linear fits was observed. In the case of TOF-SIMS its value is 0.0186 and 0.0974 for $GeCs_2^+/SiCs_2^+$ and

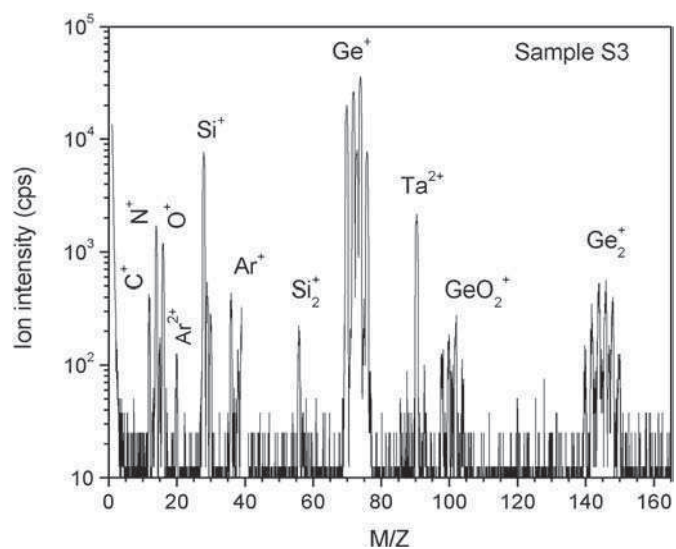

**Fig. 3.** Mass spectrum of the post-ionized sputtered neutrals measured for sample S3.

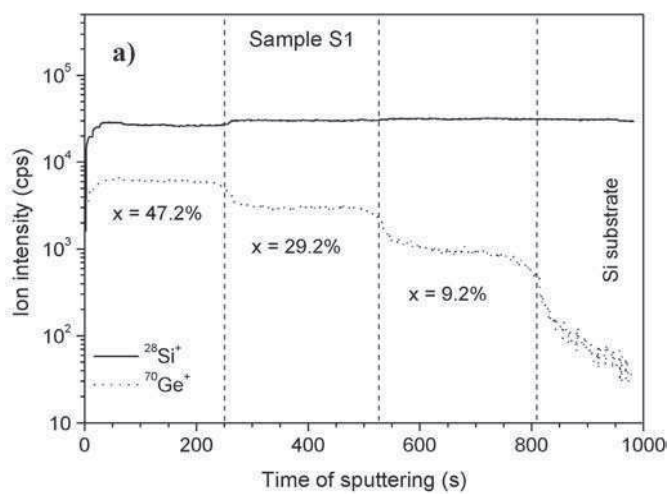

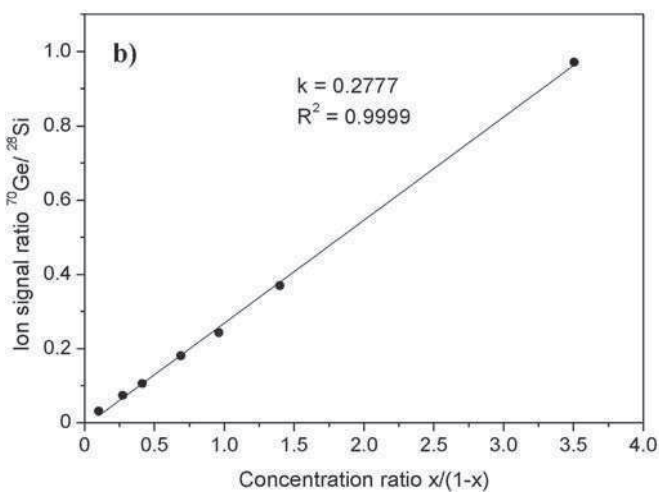

**Fig. 4.** SNMS depth profiles of the post-ionized $^{28}Si$ and $^{70}Ge$ sputtered neutrals measured for sample S1 (a), and linear fit of the calibration dependence obtained for samples S1–S3 (b), where $R^2$ is the correlation coefficient and $k$ is the calibration factor from Eqs. 2 and 3.

$^{74}$Ge/$^{30}$Si$^-$, respectively, and ($-0.0091$) for $^{70}$Ge$^+$/$^{28}$Si$^+$ in the case of SNMS. There are two main reasons of such an intercept — one is inaccuracy of the measurements, and the other is the fundamental difference in the trace calibration for the diluted samples, when Ge content is lower than 1%, and the bulk calibration of Si$_{1-x}$Ge$_x$ systems, when $x \gg 0.01$. This issue was discussed in [12], where along with Si$_{1-x}$Ge$_x$ samples ($x = 5 \sim 60\%$) the $^{70}$Ge-implanted standard verified by RBS was involved in the calibration procedure. It was found [12] that the intercept value is ($-0.0448$) in the case of the bulk calibration and zero value for the trace level of Ge, and $k$-factor is also different for the bulk and trace calibration. In [12] SIMS analyses were carried out with a magnetic sector Cameca IMS Wf using low-energy O$_2^+$ primary ions, and the values of intercept and $R^2$ were greater than in our study. We are inclined to explain the non-zero intercept value in our study mainly by inaccuracy of the measurement. However, the correctness of the bulk calibration approach for quantification of trace Ge concentration in Si$_{1-x}$Ge$_x$ systems by TOF-SIMS and SNMS needs additional experimental verification.

### 3.3. Quantitative TOF-SIMS and SNMS depth profiling of Si$_{0.63}$Ge$_{0.37}$/Si multilayer structure

TOF-SIMS depth profiles of the [10 × (12.3 nm Si$_{0.63}$Ge$_{0.37}$/34 nm Si)] structure covered by a 150 nm-thick silicon layer (S4 sample) were shown in Fig. 5. The profiles are presented in semi-logarithmic (Fig. 5a) and linear scales (Fig. 5b, only 3 first periods). The sputter time was converted into the depth of sputtering assuming a constant average sputter rate estimated via the measurements of the final crater depth after the ending of the experiments. In principle, the sputter rate of Si and Si$_{1-x}$Ge$_x$ layers should be different, however, due to a high concentration of silicon and the smaller thickness of these layers as compared with pure Si layers, such averaging does not greatly influence an accuracy of depth calibration. The degradation of the profiles with the depth of sputtering was not observed, and all peaks look symmetrical. Furthermore, the modulation of the peak intensity tends to increase towards deeper layers.

For the first SiGe layer (Fig. 5b), the full width at half of the maximum (FWHM) was found to be 10.8 nm, which is slightly lower than the value of 12.3 nm obtained by HR-XRD due to the higher measurement uncertainty of the thickness of SiGe layers as compared with TOF-SIMS [32]. Ge concentration was estimated using calibration $k$-factors obtained for the positive molecular and negative atomic secondary ions. Both approaches gave practically identical results with maximal Ge content of ca. 40 at.%, which is slightly greater than the value of 37 at.% obtained by HR-XRD.

The SNMS depth profile of the same structure is presented in Fig. 6. The depth calibration was carried out in a similar manner as TOF-SIMS. Evident profile degradation was observed. The main artifacts responsible for such degradation are the crater geometry (crater shape) and ion-induced surface roughening. In SIMS, both magnetic sector

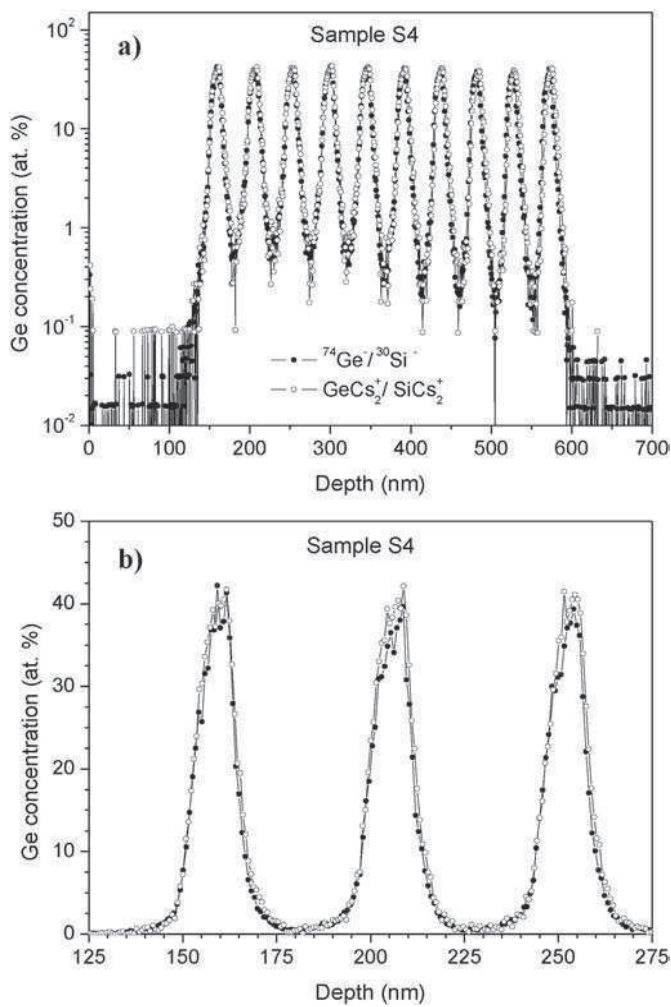

**Fig. 5.** Complete TOF-SIMS depth profiles of the [10 × (12.3 nm Si$_{0.63}$Ge$_{0.37}$/34 nm Si)] structure presented in a semi-logarithmic scale (a), and three first periods of these profiles in a linear scale (b). Quantification was carried out using preliminary obtained calibration data for the positive molecular and negative atomic secondary ions.

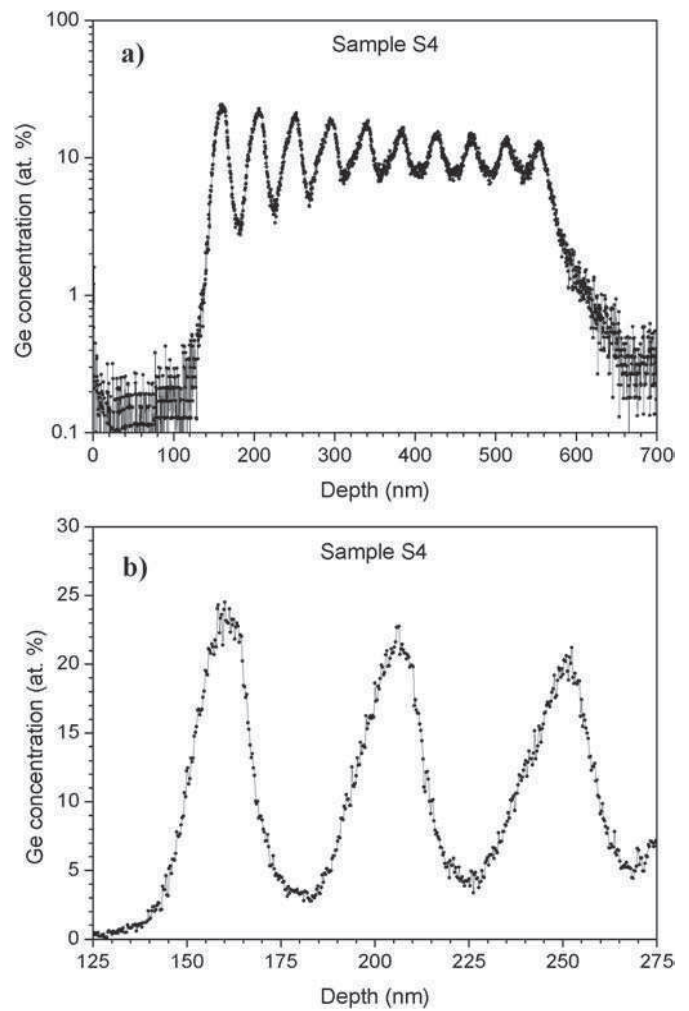

**Fig. 6.** Complete SNMS depth profile of the [10 × (12.3 nm Si$_{0.63}$Ge$_{0.37}$/34 nm Si)] structure presented in a semi-logarithmic scale (a), and three first periods of these profiles in a linear scale (b). Quantification was carried out using preliminarily obtained calibration data for the intensity ratio $^{70}$Ge$^+$/$^{28}$Si$^+$ of post-ionized sputtered neutrals.

29

and time-of flight instruments, the problems related to non-uniform crater geometry are generally overcome by limiting the analysis region to only the central part of the bottom of the crater. But, in plasma-based sputter techniques such as electron-gas SNMS, glow discharge optical emission spectroscopy (GDOES) and the novel plasma profiling TOF-MS, the whole sputter area participates in the generation of analytical signals arising from sputter atoms. The influence of crater geometry and inherent surface roughening by plasma erosion on GDOES depth profiles of $Mo/B_4C/Si$ and $Mo/Si$ multilayer samples was discussed in our previous study [33].

For the first SiGe layer (Fig. 6b), the FWHM was found to be 15 nm, that is greater than the values obtained by HR-XRD and TOF-SIMS. The peak form is asymmetrical with the length of rising edges greater than decaying ones, and this difference increases towards deeper layers. Using the calibration data, Ge concentration was estimated to be ca. 25 at.%, which is significantly lower than that obtained, by HR-XRD and TOF-SIMS

### 3.4. Reconstruction of TOF-SIMS and SNMS sputter depth profiles

Due to the high depth resolution and small ion-induced surface roughening, the degradation of the TOF-SIMS depth profiles collected for sample S4 was found negligible (Fig. 5). On the contrary, the SNMS depth profile of this sample (Fig. 6) was subjected to noticeable distortions. To understand an influence of ion-induced effects on the experimental depth profiles, we reconstructed initial Ge in-depth distributions using Hofmann's mixing-roughness-information depth model.

For the TOF-SIMS profiles, we used the following MRI parameters: the mixing length $w = 1$ nm, the surface roughness $\sigma = 2.3$ nm, and the information depth $\lambda = 0.3$ nm. The simulation was carried out in the stationary mode, i.e. all above-mentioned MRI parameters were considered constant during depth profiling. In the case of the SNMS depth profiles, we took into account nonstationary effect – the development of surface roughness of the crater bottom with sputtered depth. The roughness versus the depth of sputtering $z$ relied [34,35] on

$$\sigma(z) = \sigma_0 + a\sqrt{z}, \quad (4)$$

where $\sigma_0$ is the initial root-mean-square roughness and $a$ is the proportionality coefficient. These parameters were estimated from the experimental roughness measurements, and found to be $\sigma_0 = 2.3$ nm, $a = 1.5$. Since $w$ and $\lambda$ have a smaller impact on the depth profiling as compared with increased roughening, their values were chosen to be the same as in the case of TOF-SIMS. The segregation and diffusion of germanium were not accounted in our simulation because these are minor processes as against surface roughening.

The test model of sample S4 was introduced in the form of 10 periods of the $Si_{1-x}Ge_x$ structure with abrupt interfaces. The results of our simulation together with the experimental profiles are shown in Fig. 7a for TOF-SIMS and in Fig. 7b for SNMS. In fact, the MRI model was the same for both techniques, but roughness contribution in this model was accounted by a different manner. Namely, for simulation of the TOF-SIMS depth profiles constant roughness value was used, and a dynamically growing roughness parameter was employed in the case of the SNMS. The usage of the different MRI model parameters has allowed applying one test model for the fitting of both experimental datasets. The SiGe layer characteristics (layer's center position and FWHM) of the test model obtained by the reconstruction of the experimental profiles are presented in Table 2. The Ge content in all layers was estimated to be 40 at.%. As regards the simulated profiles shown in Fig. 7, one can see that in the case of TOF-SIMS the quality of the fitting is better than for SNMS due to the influence on the experimental SNMS depth profiles not only of the progressing surface roughening, but also of the crater effect and some other processes unaccounted in the MRI simulation.

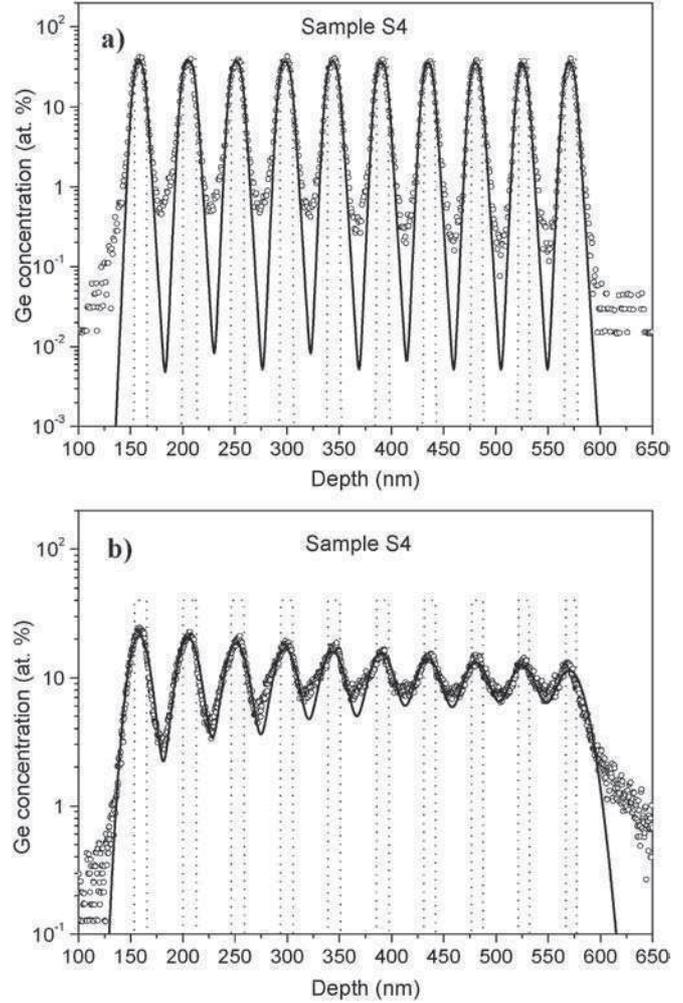

**Fig. 7.** TOF-SIMS (a) and SNMS (b) depth profiles of [10 × (12.3 nm $Si_{0.63}Ge_{0.37}$/34 nm Si)] structure. The experimental data are indicated by small circles, and the MRI simulated profiles by solid lines. Dotted lines present the reconstructed model of the structure.

**Table 2**
Characteristics of SiGe layers in the [10 × (12.3 nm $Si_{0.63}Ge_{0.37}$/34 nm Si)] structure (sample S4) obtained by the MRI reconstruction of TOF-SIMS and SNMS experimental data.

| Number of layer[a] | Center position[a] (nm) | FWHM (nm) |
|---|---|---|
| 1 | 160 | 11 |
| 2 | 207 | 13 |
| 3 | 253 | 13 |
| 4 | 300 | 12 |
| 5 | 345 | 12 |
| 6 | 392 | 12 |
| 7 | 437 | 11 |
| 8 | 482 | 11 |
| 9 | 527 | 11 |
| 10 | 572 | 11 |

[a] Layer's numeration and center position from the surface towards Si substrate.
30

## 4. Conclusions

Quantification of germanium in $Si_{1-x}Ge_x$ structures was carried out by TOF-SIMS and electron-gas SNMS. For calibration, the peak intensity ratios of $GeCs_2^+/SiCs_2^+$ and $^{74}Ge^-/^{30}Si^-$ secondary ions and post-ionized $^{70}Ge^+/^{28}Si^+$ sputtered neutrals were employed. A good linear correlation ($R^2 > 0.9997$) of these dependencies with Ge concentration for the x-value ranging from 0.092 to 0.78 was revealed for both techniques. That confirms the validity of the constant yield ratio assumption (the counterbalancing of matrix effects) in $Si_{1-x}Ge_x$ structures, especially for TOF-SIMS measurements.

The calibration data were used for quantitative depth profiling of $[10 \times (12.3\ nm\ Si_{0.63}Ge_{0.37}/34\ nm\ Si)]$ structures grown on Si (001) substrate. In the case of TOF-SIMS, the degradation of the profiles with depth of sputtering was not observed. The FWHM of the first SiGe layer was found to be 10.8 nm, and the maximal Ge content in this layer was estimated ca. 40 at.%. On the contrary, an evident degradation of the SNMS profile towards deeper layers was observed due to the influence of crater geometry and inherent surface roughening by plasma erosion. The FWHM of the first SiGe layer was found to be 15 nm, and the maximal Ge content was estimated ca. 25 at.%.

The reconstruction of sputter depth profiles of $[10 \times (12.3\ nm\ Si_{0.63}Ge_{0.37}/34\ nm\ Si)]$ structures was performed using Hofmann's mixing-roughness-information depth model in stationary and nonstationary modes for TOF-SIMS and SNMS, respectively. A good fitting of both experimental profiles was obtained using the same test model of Ge in-depth distribution and, for SNMS experiments, with proper accounting of the developing roughness. However, in the case of TOF-SIMS, the quality of the reconstruction was better than for SNMS since not only the progressing roughening, but also the crater effect and other processes unaccounted in the MRI simulation could have a significant impact on plasma sputter depth profiling.

## Acknowledgments

This work was partly supported by the Russian Foundation for Basic Research (contract No. 15-02-02947). TOF-SIMS and profilometry equipment of the Centre of Physics and Technology of Micro- and Nanostructures at IPM RAS were used for SIMS measurements, and the SNMS study was carried out at the Institute for Nuclear Research of the Hungarian Academy of Science supported by the project TÁMOP-4.2.2.A-11/1/KONV-2012-0036. We also acknowledge the Portuguese National Funding Agency FCT-MEC through the CEFITEC research grant UID/FIS/00068/2013.